# Generation and control of extreme ultraviolet free-space optical skyrmions with high harmonic generation

**Yiqi Fang,[a] Yunquan Liu,[a,b,c*]**

[a]State Key Laboratory for Mesoscopic Physics and Frontiers Science Center for Nano-optoelectronics, School of Physics, Peking University, Beijing 100871, China

[b]Collaborative Innovation Center of Extreme Optics, Shanxi University, Taiyuan, Shanxi 030006, China

[c]Beijing Academy of Quantum Information Sciences, Beijing 100193, China

**Abstract**. Optical skyrmion serves as a crucial interface between optics and topology. Recently, it has attracted great interest in linear optics. Here, we theoretically introduce a framework for the all-optical generation and control of free-space optical skyrmions in extreme ultraviolet regions via high harmonic generation. We show that by employing full Poincaré beams, the created extreme ultraviolet fields manifest as skyrmionic structures in Stokes vector fields, whose skyrmion number is relevant to harmonic orders. We reveal that the generation of skyrmionics structure is attributed to spatial-resolved spin constraint of high harmonic generation. Through qualifying the geometrical parameters of full Poincaré beams, the topological texture of extreme ultraviolet fields can be completely manipulated, generating the Bloch-type, Néel-type, anti-type, and higher-order skyrmions, etc. This work promotes the investigation of topological optics in optical highly nonlinear processes, with potential applications towards ultrafast spintronics with structured light fields.

## 1 Introduction

Skyrmions[1] are topological defects in vector fields, that can be created, moved, and annihilated. They are ubiquitous and important in condensed matter physics[2,3]. Recently, the idea of skyrmions was introduced into the communities of photonics and optics[4]. The optical analogies of skyrmions have found advanced applications in various fields, e.g., topological Hall devices[5] and deep-subwavelength microscopy[6]. Optical skyrmion can be realized when the values of electromagnetic vector fields[4,6-8] or Stokes vector fields[9-11] are taken to fully cover a unit sphere ($4\pi$) within the skyrmionic boundaries. It was first discovered that one can generate optical skyrmions in the form of surface plasmon polaritons by interacting light with nanostructures[4]. Later, the generation of free-space optical skyrmions was proposed and realized through spatial light modulation methods with digital holography[10,12] and optical focusing[13]. All these methodologies concentrate on linear optical processes. Owing to the rigid requirement of the local state of polarization for skyrmionic vector textures, the manifestation and manipulation of short-wavelength optical skyrmions are extremely rare.

High harmonic generation (HHG) is a typical highly nonlinear process in ultrafast science[14], through which the fundamental driving photons can be up-converted to a wide spectral region from vacuum ultraviolet to X-ray domain[15]. In HHG, the local states of driving laser fields are inherently resolved with sub-nanoscale spatial resolution of light-atom interaction. This nature promotes the emerging HHG studies to producing novel spatially structured light beams[16-23], which has shown great potential in sculpting extreme-ultraviolet (EUV) and soft X-ray lights. Optical skyrmions in visible light have demonstrated relevant applications in displacement sensing[24]. Looking forward, the creation of EUV optical skyrmions will offer appealing opportunities for ultra-fast and -micro metrology, since they can provide totally opposite spin state in small spatial scales, allowing a high degree of controllability when they participate directly in optical nonlinear interaction. It is highly demanding to use HHG to shape EUV fields with a polarization-dependent topological structure in a controlled manner. However, due to the electron recollision nature, the polarization control of HHG is not straightforward. The intersection between HHG and optical skyrmions have therefore remained untapped, which impedes the further exploration of topological light in EUV region.

In this work, we show how to create and control EUV free-space optical skyrmions. By employing HHG as the highly nonlinear interaction process and full Poincaré beams with different wavelengths as the driving fields, we demonstrate that the generated EUV fields reveal explicit skyrmionic structures in Stokes vectors. Their skyrmion numbers are determined by monitoring the polarity and vorticity of Stokes vector distributions, which manifest a close correction with the harmonic orders. We reveal the underlying mechanism is related to spatial-resolved spin constrain in HHG. More importantly, we further propose by employing the concept of hybrid-order Poincaré sphere pairs a novel control scenario for the topological texture of the EUV skyrmions. In this way, we show that with controlling the geometrical parameters of the driving full Poincaré beams, one can form arbitrary crucial EUV skyrmionic textures (e.g., Bloch-type, Néel-type, anti-type, and higher-order skyrmions) in HHG. This controlling strategy can be extended to other nonlinear optical processes.

## 2 Results

### 2.1 Signatures of the driving structured fields

The full Poincaré beam is a typical vector field, whose polarization state covers the full surface of Poincaré sphere. It is constructed by an arbitrary-order Laguerre-Gaussian beam and a fundamental Gaussian mode with orthogonal polarization state. Its electric field can be given by: $\mathbf{E}_{i,\omega}(\mathbf{r}_i, t) = e^{-i\omega t}[u_\omega e^{i\ell_\omega \varphi_i}(\mathbf{e}_x + i\mathbf{e}_y) + u'_\omega e^{i\ell'_\omega \varphi_i}(\mathbf{e}_x - i\mathbf{e}_y)] + c.c.$, where $t$ is time, $\omega$ is angular frequency, $\mathbf{r}_i = (\rho_i, \varphi_i)$ denotes the spatial coordinate for the incident driving fields (the notation 'i' indicates the variables in the incident plane, as seen in the inset of Fig. 1(a)), $u_\omega$ ($u'_\omega$) is the spatial complex amplitude, and $\ell_\omega$ ($\ell'_\omega$) is the topological charge ($\ell_\omega$ or $\ell'_\omega$ equals zero). Without loss of generality, $u_\omega$ ($u'_\omega$) can be approximated as a complex constant for incident laser. For conciseness, we here introduce a geometrical parameter defined as: $\eta_\omega = (\ell_\omega, \ell'_\omega, |u'_\omega|/|u_\omega|, \arg(u'_\omega) - \arg(u_\omega))$. As shown later, $\eta_\omega$ directly determines the topology of EUV fields.

To demonstrate our scheme, we start by considering one typical kind of full Poincaré beams, namely cylindrical vortex vector beams (CVVBs). We illustrate the physical scenario in Fig. 1(a), where the incident field is spatiotemporally superposed by two CVVBs with different wavelengths (800-nm + 400-nm). The geometrical parameters of the 800-nm and 400-nm CVVBs are given by $\eta_{\omega 1} = (-2, 0, 1, 0)$ and $\eta_{\omega 2} = (0, 2, 1, 0)$. The notations $\omega_1$ and $\omega_2$ represent the angular frequency of 800-nm and 400-nm, respectively. As shown in Fig. 1(a), for each CVVB, it carries both radial polarization state (i.e., optical polarization singularity) and spiral phase (optical phase singularity).

In HHG, atoms directly interact with the focal field of incident laser. It is important to know the signatures of focal field distribution. We therefore simulate the focal field of CVVB by using Richards-Wolf vectorial diffraction method[25]. We use 20th order Gauss–Legendre integral formula to numerically calculate the integral, and the grid size was chosen as 50 μm in the focal plane with a spacing of 0.05 μm. In our simulation, the numerical aperture is selected to be NA = 0.02. It is a typical focusing condition of HHG experiments. In this way, we can obtain the focal intensity distributions of both 800-nm and 400-nm field [up panel in Fig. 1(b)]. One can notice that the focal intensity distributions manifest as Gaussian-like structures, which differ from the typical donut-shaped structure of individual phase or polarization singularity. Such exotic distribution largely enriches the highly nonlinear interaction in spatial domain. Furthermore, we calculate the normalized spin angular momentum (SAM) density of focal fields[26,27]. As shown in the bottom

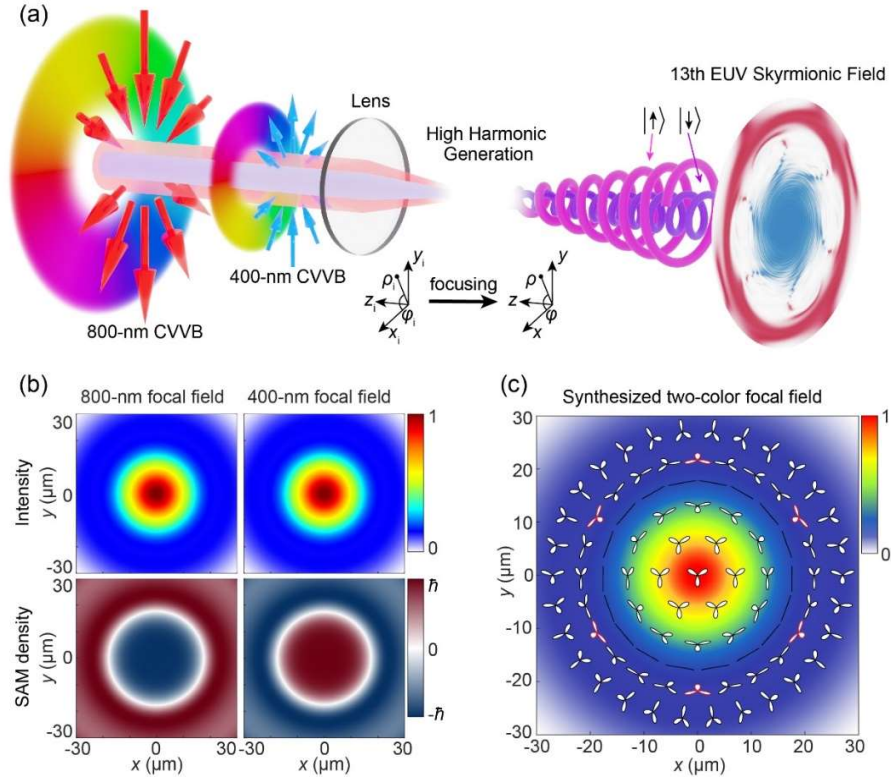

**Fig. 1** (a) Schematic of generating EUV skyrmions with HHG. A synthesized two-color (800-nm + 400-nm) CVVB field is employed to generate HHG with non-trivial spatial spin distribution. The red and blue arrows represent the instantaneous electric field vectors of 800- and 400-nm incident fields, respectively, and the accompanied pseudocolor maps show their angular phase structures. We control the beam width of the incident 800-nm field is twice that of 400-nm field, so that the focal distributions of intensities of 800-nm and 400-nm CVVBs are the same. We note here that if they are not the same, one cannot generate the skyrmions presented in this work. Here, we show the EUV Skyrmionic fields can be generated, and the 13th harmonic spins down at the beam center and spins up at the surrounding. The inset illustrates the coordinates in incident plane and focal plane, respectively. (b) The distributions of intensity (up panel) and normalized *z*-direction SAM density (bottom panel) of 800-nm and 400-nm CVVB in the focal plane individually. (c) The focal electric field structure of the two-color synthesized CVVB. The pseudocolor map indicates the intensity distribution of the driving beam. The overlapped Lissajous figures illustrate the local polarization states, in which the red profiles show the 6-fold spatial symmetry of the two-color synthesized field.

panel of Fig. 1(b), spatial-varying polarization is generated. It is interesting to find the presence of circular polarizations in focal plane despite the incident light being linearly polarized throughout. Here, we emphasize that both the peculiar intensity distribution and SAM distribution of focal field originate from spin-orbit interaction of light[28]. The underlying origin of such spin-orbit interaction comes from the time-varying polarization distribution in the wavefront of incident CVVB[28,29]. When the 800-nm and 400-nm CVVBs are spatiotemporally synthesized at focus, three-leaf polarizations show up [Fig. 1(c)]. This polarization has played an important role in the studies of HHG, which allows to emit circularly polarized EUV light in HHG[30]. Most of the relevant works only involve laser field with a spatially homogeneous three-leaf polarization. In our case, this two-color synthesized field, known as a Lissajous beam[31,32] carries a versatile polarization state, and the polarization varies rapidly in space.

## 2.2 Generation of EUV skyrmions with CVVBs

To explore the HHG driven by the synthesized fields, we perform numerical simulation by solving Schrödinger equation for Hydrogen within strong-field approximation (SFA)[33]. The high harmonic field with angular frequency $\omega$ was calculated from the Fourier components of dipole moment, $\mathbf{P}(\mathbf{r}, \omega) = \omega^2 \int \mathbf{D}(\mathbf{r}, t) e^{-i\omega t} dt$. Here, the dipole moment depends on both space($\mathbf{r}$) and time($t$), given by $\mathbf{D}(\mathbf{r},t) = -i \int_{-\infty}^{t} dt' [-2\pi i/(t-t'-i\delta)]^{3/2} \mathbf{d}^*[\mathbf{p}_s+\mathbf{A}(\mathbf{r},t)] \cdot \mathbf{E}(\mathbf{r},t') \cdot \mathbf{d}[\mathbf{p}_s+\mathbf{A}(\mathbf{r},t')] \exp[-iS(\mathbf{r},\mathbf{p}_s,t)] + c.c.$, in which $\delta$ is an arbitrary small positive constant, $\mathbf{p}_s$ is saddle point momentum calculated by saddle point equation $\nabla_\mathbf{p} S(\mathbf{r},\mathbf{p}_s,t) = 0$, $S(\mathbf{r},\mathbf{p}_s,t) = \int_{t'}^{t} dt'' \{[\mathbf{p}+\mathbf{A}(\mathbf{r},t'')]^2/2+I_p\}$ is the quasi-classical action, $d^*[.]$ and $d[.]$ contribute to photoelectron recombination and ionization process, $\mathbf{E}$ is the focal electric field obtained by Richards-Wolf vectorial diffraction method and $\mathbf{A}$ is the vector potential of light. We insert the focal electric field into the expression of HHG dipole moment and obtain the spatially resolved HHG field of different orders by changing the harmonic frequency $\omega$.

In order to elucidate the advantages of two-color scenario in generating EUV skyrmions, we compare the HHG driven by three different fields, i.e., the 800-nm CVVB, 400-nm CVVB and their synthesized two-color light field. In simulation, the peak intensities of each driving fields are taken to be $I = 5 \times 10^{14}$ W/cm$^2$, and their temporal envelops are selected to be a sin$^2$ function of $\tau =$ 21.3-fs duration time. In Figs. 2(a)-2(c), we show the HHG spectra for different drivers. Due to the joint restriction taken by parity, energy, and momentum conservations in HHG, harmonic spectra reveal the orders of $(2n-1)$th, $(4n-2)$th, and $n$th for the three different drivers, respectively.

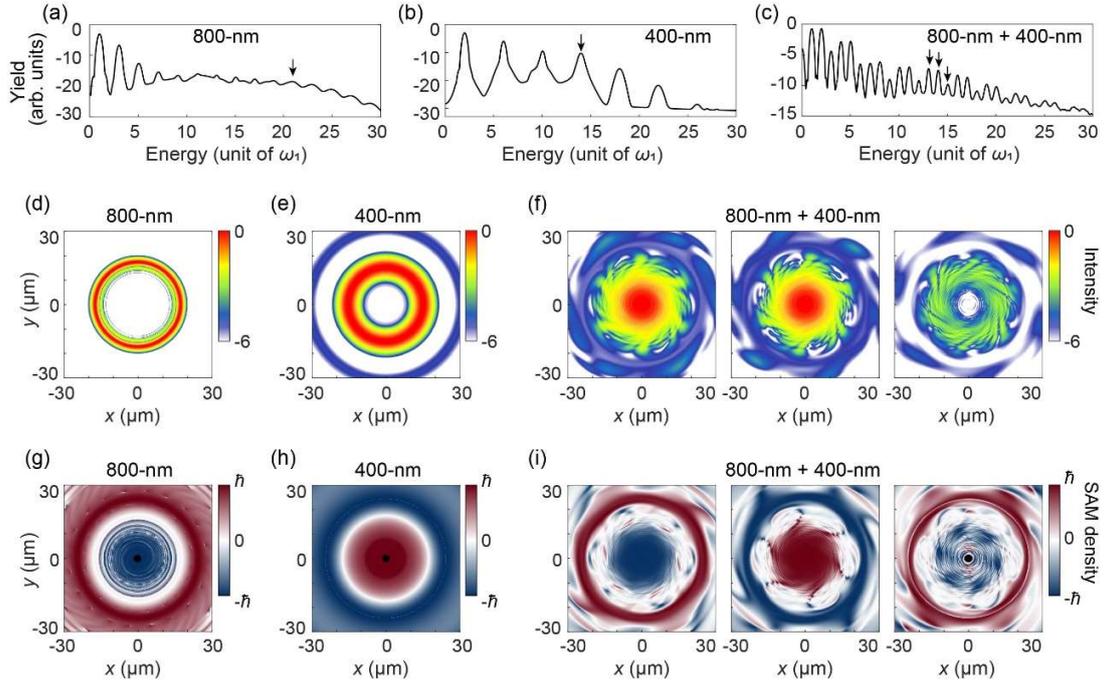

**Fig. 2** Spatio-spectral signatures of HHG driven by the CVVBs. (a)-(c) High harmonic spectra. (d)-(f) Spatial intensity distributions of logarithmic scale. (g)-(i) Spatial distribution of normalized SAM density along *z*-direction. The HHG drivers are the 800-nm CVVB (a,d,g), 400-nm CVVB (b,e,h), and their synthesized two-color CVVB (c,f,i), respectively. The harmonic orders marked by black arrows in (a)-(c) correspond to each column in (d)-(i).

In time domain, the generated field in HHG is spatially structured attosecond pulse trains. To identify the real-space topology of the harmonics, we confine one specific frequency of harmonics and extract its spatial intensity distributions [Figs. 2(d)-2(f)]. It can be noticed that when the driver is a single-color CVVB, the harmonics manifest as annulus structures. By contrast, the intensity of harmonics with the two-color driving field are different, which reveals a Gaussian-like structure for $(3n+1)$th and $(3n-1)$th harmonic or annulus structure for $3n$th harmonic. Due to the peculiar polarization texture of the driving two-color beam, the $3n$th harmonic has significantly lower yield than the $(3n\pm1)$th harmonics. In Figs. 2(g)-2(i), we show the corresponding SAM distributions. For each harmonic, the SAM density is spatially varied. Upon closer inspection, in the case of two-color CVVB driver, the SAM distribution of produced harmonics reveals a 6-fold symmetrical distribution [Fig. 2(i)]. This unique distribution can be attributed to the inherent 6-fold symmetry of the driving field, which is illustrated by the red profiles depicted in Fig. 1(c).

To form a skyrmionic field, local SAM should orient oppositely at the center of field compared to the surrounding. Our simulation indicates that one cannot obtain an explicit EUV skyrmionic field with single-color CVVB drivers in HHG. This is because the recollision of electrons is rigidly forbidden in a single-color circularly polarized local field[15], resulting in an extremely low photon yield at the center of the harmonic beam [as shown in Figs. 2(d) and 2(e)]. Consequently, the spatial spin reversal is not complete. By contrast, such obstacles can be overcome in the $(3n\pm1)$th harmonics with a two-color driver. In this case, the yield at the center is substantial, and the SAM is oriented oppositely at the center compared to the surrounding, leading to a non-trivial real-space topology.

To prove that the created $(3n\pm1)$th harmonics from two-color CVVB fields are skyrmions, we then analyze their spatial vector texture. Conventionally, skyrmions are characterized by skyrmion number $Q$, defined as[34]: $Q = (1/4\pi)\cdot\iint_{\rho_0}\mathbf{S}\cdot(\partial_x\mathbf{S}\times \partial_y\mathbf{S})dxdy$, where $\mathbf{S}(x, y)$ and $\rho \leq \rho_0$ denote the vector fields and the defined region for constructing skyrmions in $(x, y)$ plane. The vector fields of a two-dimensional skyrmion can be written as $\mathbf{S}(\rho\cos(\varphi),\rho\sin(\varphi))=(\cos[\alpha(\varphi)]\sin[\beta(\rho)],\sin[\alpha(\varphi)]\sin[\beta(\rho)],\cos[\beta(\rho)])$, where $\rho$ and $\varphi$ are polar coordinate in focal plane, and $\alpha$ and $\beta$ are spherical coordinates in Poincaré sphere, as shown in Fig. 3(a). In this way, $Q$ reads:

$$Q = \frac{1}{2}[\cos\beta(\rho)]_{\rho=0}^{\rho=\rho_0} \cdot \frac{1}{2\pi}[\alpha(\varphi)]_{\varphi=0}^{\varphi=2\pi} = q\cdot m. \quad (1)$$

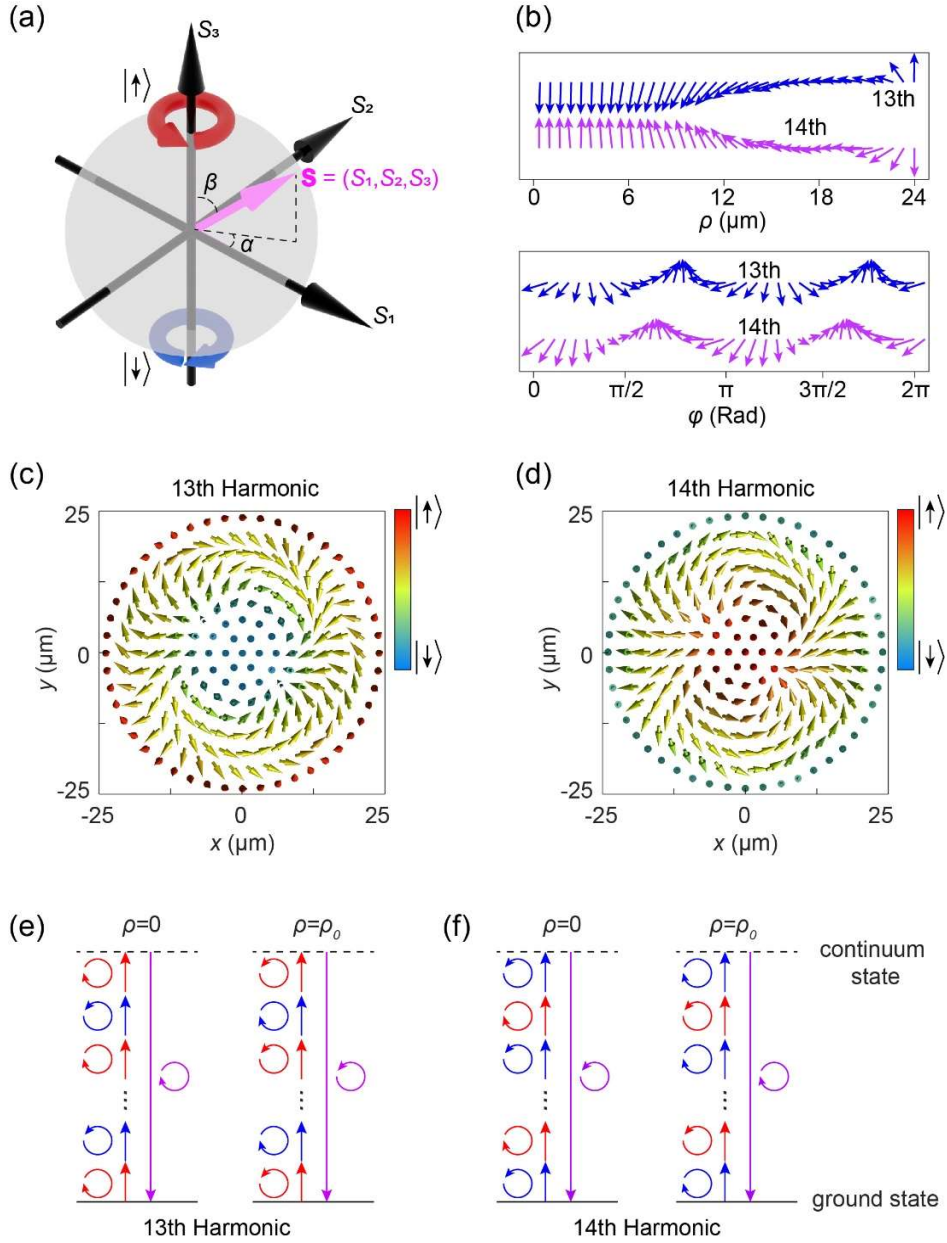

**Fig. 3** Characterization of the EUV skyrmions in HHG. (a) Unit Poincaré sphere defined by the normalized Stokes parameters, **S** = ($S_1$, $S_2$, $S_3$). At the north and south poles of Poincaré sphere, polarization is spin-up and spin-down, respectively. (b) The projection of Stokes parameters of 13th and 14th harmonics in the $S_1$-$S_3$ plane as a function of radial distance (upper panel), and the projection in the $S_1$-$S_2$ plane as a function of azimuthal angle (lower panel). (c), (d) Spatial distributions of Stokes vectors for the 13th and 14th harmonics in the focal plane, in which the color scheme of Stokes vector is encoded by the value of $S_3$. (e), (f) Physical picture of generating skyrmions in HHG.

The skyrmion number is divided into two integers, $q$ and $m$, in which $q$ describes the polarity of skyrmions and $m$ indicates the vorticity of skyrmions, respectively. In order to create a skyrmionic field, $q$ and $m$ should both be non-zero. That is, the field vector has to be reversed comparing the center $\rho = 0$ with its confinement $\rho = \rho_0$, and meanwhile it should also cover the unit Poincaré sphere as $\rho$ increases from $\rho = 0$ to $\rho = \rho_0$.

In our configuration, the focal light field can be approximated as a paraxial beam, and thus we explore the possibility of generating EUV field with a skyrmionic structure in Stokes vector fields, $\mathbf{S} = (S_1, S_2, S_3)$. Each parameter of $\mathbf{S}$ corresponds to an expectation value of a Pauli matrix for a photonic spin state. The skyrmionic boundary is here selected to be $\rho_0 = 24$ μm, in which the focal 800-nm and 400-nm driving CVVBs are circularly polarized. We then inspect the topological character of the skyrmions by the stereographic projection of Stokes vectors [Figs. 3(b)-(d)]. One can notice that the Stokes vectors point to the opposite direction at $\rho = 0$ and $\rho = \rho_0$, and their orientations change in a vortex-type [Figs. 3(c) and 3(d)]. To obtain the specific topological charge, $Q$, we inspect the azimuthal and radial variation of the Stokes vector distribution [Fig. 3(b)]. It can be seen the direction of Stokes vector is changed from downward (upward) to upward (downward) for the 13th (14th) harmonic. This indicates the polarity of the 13th and 14th harmonics is given by $q_{13} = -1$ and $q_{14} = 1$, respectively. Besides, their transverse components rotate twice counterclockwise from $\varphi = 0$ to $\varphi = 2\pi$, meaning the vorticity of skyrmion is given by $m_{13} = m_{14} = 2$. Taking both the polarity and vorticity of skyrmions into account, the skyrmion numbers can be determined to be $Q_{13} = q_{13} \cdot m_{13} = -2$ and $Q_{14} = q_{14} \cdot m_{14} = 2$ for these two high harmonics, respectively. Such two EUV skyrmionic fields can be categorized as two-order skyrmions[3].

It is interesting to note that the skyrmion number is dependent on harmonic order. In the picture of absorbing photons, the allowed photon absorption channel for specific harmonics can be written by $(n_{\omega 1}, n_{\omega 2})$, in which $n_{\omega 1}$ and $n_{\omega 2}$ denote for the number of photons absorbed from 800-nm and 400-nm CVVBs. Since the driving field is spatially structured, the allowed photon channel is spatially varied. Although, the allowed photon absorption channels cannot be solely determined at the positions in $0 < \rho < \rho_0$, but the channel is well known at $\rho = 0$ and $\rho = \rho_0$. There, the channel is given by (5, 4) for the 13th harmonic. That is, the number of absorbed 800-nm photons is one more than the number of 400-nm. Therefore, the local spin states at $\rho = 0$ and $\rho = \rho_0$ are the same as that of 800-nm photon for 13th harmonic [Fig. 3(e)]. Similarly, the corresponding photon channel of 14th harmonic is (4, 5), and thus the local spin is the same as 400-nm photon at $\rho = 0$ and $\rho = \rho_0$ [Fig. 3(f)]. As indicated in Fig. 1(b), the local spin of 800-nm and 400-nm CVVBs are inverse, leading to the opposite polarities of the 13th and 14th EUV skyrmions. On the other hand, the vorticity of EUV skyrmions is influenced by the focal phase distribution of the 800-nm and 400-nm CVVBs, which is independent of the harmonic order. Hence, the vorticities are the same for 13th and 14th skyrmions. As a result, the skyrmion numbers of 13th and 14th skyrmions are opposite.

*2.3 Complete control over the texture of EUV skyrmions.*

The behaviors and stabilization of skyrmions highly depend on their topological texture, and thus we then demonstrate how to control the topology of EUV skyrmions. We extend the driving fields from CVVBs to general full Poincaré beams, and we also resort to the synthetization of two-color (800-nm and 400-nm) beams. To this end, we start by giving an analytical expression for the focal field of full Poincaré beams as

$$\mathbf{E}(\mathbf{r},t) = e^{-i\omega t} C \left[ u U_\ell e^{i\ell\varphi} (\mathbf{e}_x + i\mathbf{e}_y) + u' U_{\ell'} e^{i\ell'\varphi} (\mathbf{e}_x - i\mathbf{e}_y) \right] + c.c.. \quad (2)$$

In Eq. (2), $C = i^\ell kf/2$ ($k$ being wavenumber, $f$ being focal distance) and $U_\ell$ are introduced by optical focusing. $U_\ell$ modulates the complex amplitude of focal field, and it is expressed as

$$U_\ell = \int_0^{\theta_m} \cos^{\frac{1}{2}}\theta \sin\theta (\cos\theta + 1) J_\ell(\zeta) d\theta, \quad (3)$$

where $J_\ell$ stands for the $\ell$-order Bessel function of the first kind, and $\theta$ is polar angle in the output pupil whose maximum value, $\theta_m$, is determined by numerical aperture (NA) of lens. Here, it should be noted that for a full Poincaré beam, its topology is largely determined by the phase difference between its two orthogonal components, $\mathbf{e}_x + i\mathbf{e}_y$ and $\mathbf{e}_x - i\mathbf{e}_y$. According to Eq. (3), the sign of $U_\ell$ endows additional phase difference between these two components beyond the initial one, $\arg(u)$ $-\arg(u')$. It implies that the focusing process itself modulates the topology of the laser field, exerting a significant influence on the topology of EUV skyrmions in HHG.

With Eq. (2), we then propose two empirical restriction rules for the geometrical parameters $\eta$ of 800-nm and 400-nm incident full Poincaré beams, which ensure the generated EUV fields in HHG exhibit skyrmionic characteristics. (i), considering the spin constraints in HHG, the local SAMs of 800- and 400-nm full Poincaré beams should have opposite orientations in the focal plane. It requires that the third components of their Stokes vectors ($S_{3,\omega 1}$ and $S_{3,\omega 2}$) exhibit opposite signs, leading to

$$\ell_{\omega 1} = -\ell'_{\omega 2},\ \ell'_{\omega 1} = -\ell_{\omega 2},\ \text{and}\ |u'_{\omega 1}|/|u_{\omega 1}| = |u_{\omega 2}|/|u'_{\omega 2}|. \tag{4}$$

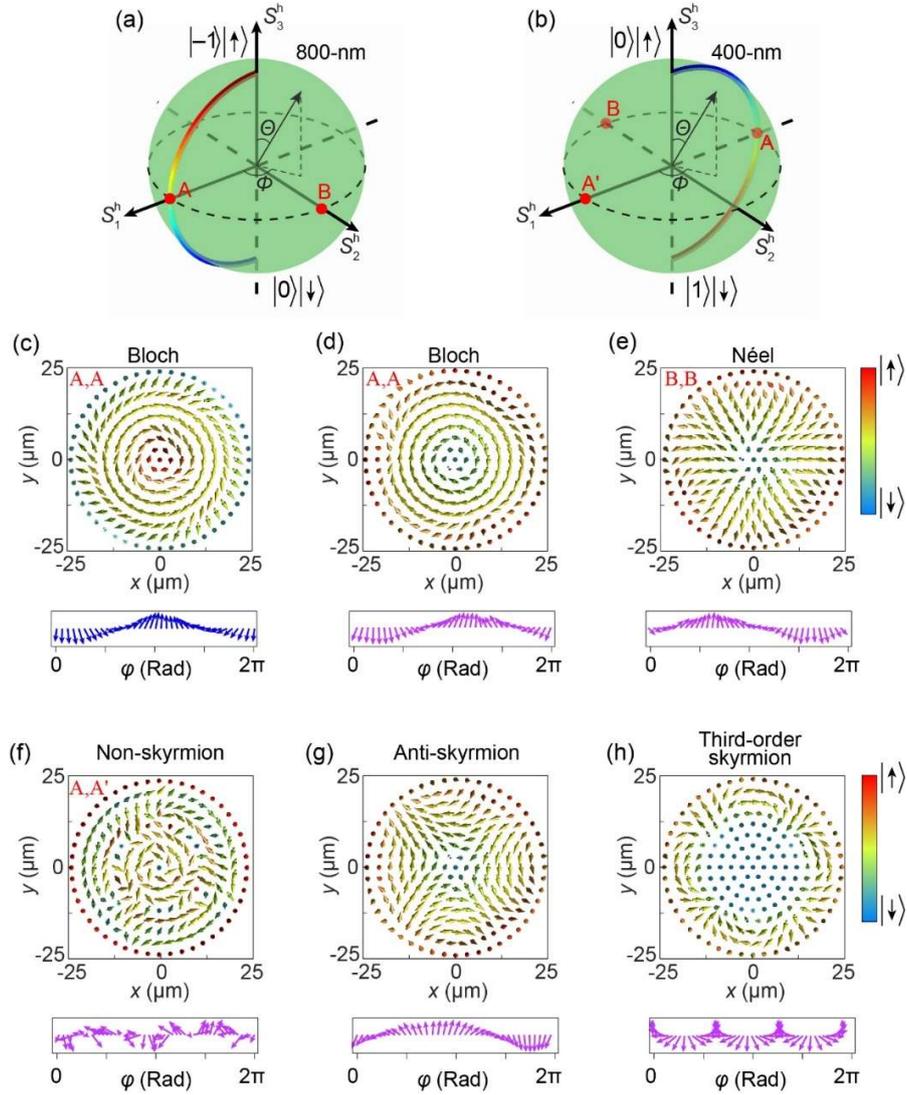

**Fig. 4** Control over the topological texture of EUV skyrmions. (a), (b) HyOPSP formed by an 800-nm Poincaré beam ($\ell_{\omega 1} = -1$ and $\ell'_{\omega 1} = 0$) and a 400-nm Poincaré beam ($\ell_{\omega 2} = 0$ and $\ell'_{\omega 2} = 1$). The varying colors on the spheres are used to show the qualified combination of two-color Poincaré beams. Points A and B on the HyOPSP indicate the positions of two-color Poincaré beams, which can generate (c), (d) Bloch-type EUV skyrmion and (e) Neel-type EUV skyrmion in HHG. (f) If selecting the point A in 800-nm sphere and A' in 400-nm sphere, the generated EUV field reveals a non-skyrmion structure. (g), (h) Formation of anti-skyrmion, and third-order skyrmion, respectively. In each diagram, lower panel shows the projection of Stokes vectors in the $S_1$-$S_2$ plane as a function of azimuthal angle. The harmonic orders are 13th for (c) and 14th for (d-f).

(ii), the transverse components of Stokes vectors (i.e., $S_1$ and $S_2$) determine the direction of long-axis of the focal electric field ellipse. To create regular spatial polarization distributions for harmonics, $S_{1,\omega 1}$ and $S_{2,\omega 1}$ need to be the same as $S_{1,\omega 2}$ and $S_{2,\omega 2}$, respectively. According to Eq. (2), $S_1$ and $S_2$ are determined by the relative phase between right- and left-handed components, i.e., $\arg(uU_\ell) - \arg(u'U_\ell)$. Thus, the two-color full Poincaré beams are further confined by

$$\frac{\pi}{2}(\kappa_{\omega 1} - \kappa'_{\omega 1}) - [\arg(\mu'_{\omega 1}) - \arg(\mu_{\omega 1})] = \frac{\pi}{2}(\kappa_{\omega 2} - \kappa'_{\omega 2}) - [\arg(\mu'_{\omega 2}) - \arg(\mu_{\omega 2})], \tag{5}$$

in which $\kappa = [\text{sign}(\ell)]^{|\ell|}$ (sign($\ell$) = 1 for $\ell \geq 0$ and sign($\ell$) = $-1$ for $\ell < 0$) is related to the additional phase introduced by $U_\ell$, and $\arg(u') - \arg(u)$ is related with the initial phase of incident field.

Confined by the above rules, for each 800-nm full Poincaré beam, there is only one qualified 400-nm full Poincaré beam. Taking an 800-nm full Poincaré beam with $\ell_{\omega 1} = -1$ and $\ell'_{\omega 1} = 0$ as an example, the topological charges of a qualified 400-nm beam should be $\ell_{\omega 2} = 0$ and $\ell'_{\omega 2} = 1$. Then, to visualize our controlling strategy of EUV skyrmions, we employ the concept of hybrid-order Poincaré sphere pair (HyOPSP), which is comprised of two hybrid-order Poincaré spheres[35] at different wavelengths. Each point on the sphere represents one specific full Poincaré beam. In Figs. 4(a) and 4(b), we display the corresponding HyOPSP, in which $S_1^h$, $S_2^h$, and $S_3^h$ are the redefined Stokes parameters serving as the sphere's Cartesian coordinates. $\Phi = \tan^{-1}(S_2^h/S_1^h) = \arg(u) - \arg(u')$ and $\Theta = \cos^{-1}(S_3^h)$ are the azimuthal angle and pitching angle, respectively. We colorize some points on these two spheres ($\Phi = 0$, $\Theta \in [0, \pi]$ for 800-nm and $\Phi = \pi$, $\Theta \in [0, \pi]$ for 400-nm,). The points in the same colors construct the qualified two-color full Poincaré beams which can generate EUV skyrmions in HHG. For each point on the 800-nm hybrid-order Poincaré sphere, one can find a unique point on 400-nm sphere that meets the above restriction rules [equation (4) and equation (5)].

Then we select some typical points on the HyOPSP to demonstrate the control over topological textures of skyrmions. At points A, the geometrical parameters of 800- and 400-nm full Poincaré beams are $\eta_{\omega 1} = (-1, 0, 1, 0)$ and $\eta_{\omega 2} = (0, 1, 1, \pi)$, respectively. As shown in Figs. 4(c) and 4(d), the HHG simulations demonstrate that the generated EUV fields manifest as Bloch-type skyrmions, whose skyrmion numbers are $Q = 1$ ($q = 1$, $m = 1$) for 13th harmonic and $Q = -1$ ($q = $

−1, $m = 1$) for 14th harmonic. Through changing the positions on the HyOPSP, the topological texture can be adjusted. At the points B, the geometrical parameters are given by $\eta_{\omega 1} = (-1, 0, 1, \pi/2)$ and $\eta_{\omega 2} = (0, 1, 1, 3\pi/2)$, and the resulted harmonic field reveals as the Néel-type skyrmions of $Q = -1$ ($q = -1$, $m = 1$) [Fig. 4(e)]. Besides the Bloch-type and Neel-type skyrmions, one can also shape the EUV skyrmions with arbitrary topological textures, such as anti-skyrmion of $Q = 1$ ($q = -1$, $m = -1$) [$\eta_{\omega 1} = (1, 0, 1, 0)$ and $\eta_{\omega 2} = (0, -1, 1, \pi)$] [Fig. 4(g)] and third-order skyrmion of $Q = -3$ ($q = -1$, $m = 3$) [$\eta_{\omega 1} = (-3, 0, 1, 0)$ and $\eta_{\omega 2} = (0, 3, 1, \pi)$] [Fig. 4(h)]. One can notice the topology of EUV skyrmions in HHG is closely related with the geometrical parameters of incident laser. The first two parameters of $\eta$ control the polarity and vorticity of EUV skyrmions, while the last two parameters of $\eta$ influence the skyrmionic texture (such as Néel- type and Bloch-type).

For comparison, if we select two-color full Poincaré beams that does not satisfy the restriction rules of $\eta$, e.g., point A on the 800-nm sphere [$\eta_{\omega 1} = (-1, 0, 1, 0)$] and point A' on the 400-nm sphere [$\eta_{\omega 2} = (0, 1, 1, 0)$], one can see the formed Stokes vectors of the EUV field distributes desultorily and the HHG field is not a skyrmion [Fig. 4(f)]. Overall, the texture of EUV skyrmions can be controlled by using the driving two-color full Poincaré beams with qualified geometrical parameters. Such a robust methodology shows its uniqueness in free-space optical skyrmions, which has never been unveiled for the optical skyrmions in guided modes or surface plasmon.

## 3 Discussion and Conclusion

This work demonstrates the generation and control of EUV skyrmions. On the one hand, the generation of EUV skyrmions roots in the transfer of photon angular momentum between driving fields and high harmonics. From the view of electron motion, the transfer of photon orbital angular momentum is realized by the spatially resolved recombination time of electrons in the interaction region. As for the transfer of photon SAM, it is first recorded by the ionized electron trajectory (or electron orbital angular momentum) with respect to nucleus and then taken by high harmonics at the instant of recombination. It should be noted that the spin-orbit interaction of light is critical in modifying the distribution of photon angular momentum in focal plane, giving rise to the generation of EUV skyrmions in HHG. On the other hand, the control of EUV skyrmions is realized by adjusting the spatial mode of two-color driving fields. In principle, the selection of two-color laser wavelength affects the spectrum of HHG, but it has little influence on the formation and control of EUV skyrmions. Compared with single-color scenario, the two-color scenario has significant advantages. It introduces more degrees of freedom that allows one to finely control the spatial structure of EUV skyrmionic fields. Moreover, two-color fields enable the creation of bright circularly polarized HHG, where the spin states of driving fields can be imprinted onto different orders of harmonics. Hence, one can sculpt the spatial mode of 3-dimensional stokes parameters for emitted EUV fields (Fig. 4c-h). Here, the employment of HyOPSP presents a clear picture for the configuration of two-color full Poincaré, and it can be easily expanded to many fields such as four-wave mixing.

To summarize, we have shown theoretically that HHG makes it possible to generate and manipulate the EUV free-space optical skyrmions. The robust control methodology shows its uniqueness in EUV optical skyrmions, which has never been unveiled for the optical skyrmions in surface plasmon polaritons or guided modes. This work presents an important interface between strong-field physics and topology[36,37]. The generation and control of EUV skyrmions open a door to access the skyrmionic dynamics in the broad ultraviolet region. Magnetic skyrmion Hall effect has been experimentally discovered recently[38], however, to our knowledge, its optical analog has been only studied in theory[6]. Since the EUV optical skyrmions reveals a regularly spatial distribution of photon spin-orbital state, its highly nonlinear process and ability in ultrahigh spatial resolution may enable the observation of optical skyrmion Hall effect. Looking broadly, the EUV skyrmions endow one laser beam with totally opposite chiral responses when interacting with chiral matters, and thus it could offer new opportunities in the studies of laser-based spatial separation of enantiomers via photoionization. Furthermore, due to the elaborate spatial polarization structure of EUV skyrmion, it also has implications in HHG-induced EUV nanoscale imaging[39].

*Code, Data, and Materials Availability*

Data underlying the results presented in this paper may be obtained from the authors upon reasonable request.

*Acknowledgments*

We thank the finance support by the National Key R&D Program (No. 2022YFA1604301) and National Science Foundation of China (Grant No. 92050201 and 11774013). The authors declare no conflicts of interest.

*References*

1. T. H. R. Skyrme, "A unified field theory of mesons and baryons," *Nucl. Phys.* **31**, 556-569 (1962).

2. Y. Tokura, and N. Kanazawa, "Magnetic skyrmion materials," *Chem. Rev.* **121**(5), 2857-2897 (2021).

3. B. Göbel, I. Mertig, and O. A. Tretiakov, "Beyond skyrmions: Review and perspectives of alternative magnetic quasiparticles," *Physics Reports* **895**, 1–28 (2021).

4. S. Tsesses, E. Ostrovsky, K. Cohen, B. Gjonaj, N. H. Lindner, and G. Bartal, "Optical skyrmion lattice in evanescent electromagnetic fields," *Science* **361**(6406), 993-996 (2018).

5. A. Karnieli, S. Tsesses, G. Bartal, and A. Arie, "Emulating spin transport with nonlinear optics, from high-order skyrmions to the topological Hall effect," *Nat. Commun.* **12**(1), 1092 (2021).

6. L. Du, A. Yang, A. V. Zayats, and X. Yuan, "Deep-subwavelength features of photonic skyrmions in a confined electromagnetic field with orbital angular momentum," *Nat. Phys.* **15**(7), 650−654 (2019).

7. T. J. Davis, et al., "Ultrafast vector imaging of plasmonic skyrmion dynamics with deep subwavelength resolution," *Science* **368**(6489), eaba6415 (2020).

8. Y. Shen, Y. Hou, N. Papasimakis, and N. I. Zheludev, "Supertoroidal light pulses as electromagnetic skyrmions propagating in free space," *Nat. Commun.* **12**(1), 5891 (2021).

9. S. Gao, F. C. Speirits, F. Castellucci, S. Franke-Arnold, S. M. Barnett, and J. B. Götte, "Paraxial skyrmionic beams," *Phys. Rev. A* **102**(5), 053513 (2020).

10. Y. J. Shen, E. C. Martínez, and C. Rosales-Guzmán, "Generation of optical skyrmions with tunable topological textures" *ACS Photonics* **9**(1), 296–303 (2022).

11. W. Lin, Y. Ota, Y. Arakawa, and S. Iwamoto, "Microcavity-based generation of full Poincaré beams with arbitrary skyrmion numbers," *Phys. Rev. Research* **3**(2), 023055 (2021).

12. Y. Shen, B. Yu, H. Wu, C. Li, Z. Zhu, & A. V. Zayats, "Topological transformation and free-space transport of photonic hopfions," *Advanced Photonics*, **5**(1), 015001 (2023).

13. R. Gutiérrez-Cuevas and E. Pisanty "Optical polarization skyrmionic fields in free space," *J. Opt.* **23**(2), 024004 (2021).

14. P. B. Corkum, "Plasma perspective on strong field multiphoton ionization," *Phys. Rev. Lett.* **71**(13), 1994 (1993).

15. T. Popmintchev, et al., "Bright coherent ultrahigh harmonics in the keV x-ray regime from mid-infrared femtosecond lasers," *Science* **336**(6086), 1287-1291 (2012).

16. J. Wätzel, and J. Berakdar, "Topological light fields for highly non-linear charge quantum dynamics and high harmonic generation," *Opt. Express* **28**(13), 19469-19481 (2020).

17. M. Zurch, C. Kern, P. Hansinger, A. Dreischuh, and C. Spielmann, "Strong-field physics with singular light beams," *Nat. Phys.* **8**(10), 743-746 (2012).

18. K. M. Dorney et al., "Controlling the polarization and vortex charge of attosecond high-harmonic beams via simultaneous spin–orbit momentum conservation," *Nat. Photonics* **13**(2), 123-130 (2019).

19. L. Rego et al., "Generation of extreme-ultraviolet beams with time-varying orbital angular momentum," *Science* **364**(6447), eaaw9486 (2019).

20. F. Kong et al., "Controlling the orbital angular momentum of high harmonic vortices," *Nat. Commun.* **8**(1), 14970 (2017).

21. D. Gauthier et al., "Tunable orbital angular momentum in high-harmonic generation," *Nat. Commun.* **8**(1), 14971 (2017).

22. L. Rego, J. San Román, A. Picón, L. Plaja, and C. Hernández-García, "Nonperturbative twist in the generation of extreme-ultraviolet vortex beams," *Phys. Rev. Lett.* **117**(16), 163202 (2016).

23. Y. Fang, S. Lu, and Y. Liu, "Controlling photon transverse orbital angular momentum in high harmonic generation," *Phys. Rev. Lett.* **127**(27), 273901 (2021).

24. A. Yang, X. Lei, P. Shi, F. Meng, M. Lin, L. Du, and X. Yuan, "Spin-Manipulated Photonic Skyrmion-Pair for Pico-Metric Displacement Sensing". *Advanced Science* **10**(12), 2205249 (2023).

25. K. S. Youngworth, T. G. Brown, "Focusing of high numerical aperture cylindrical-vector beams," *Opt. Express* **7**(2), 77–87 (2000).

26. K. Y. Bliokh, A. Y. Bekshaev, and F. Nori, "Dual electromagnetism: helicity, spin, momentum and angular momentum," *New J. Phys.* **15**(3), 033026 (2013).

27. K. Y. Bliokh, J. Dressel, and F. Nori, "Conservation of the spin and orbital angular momenta in electromagnetism," *New J. Phys.* **16**(9), 093037 (2014).

28. Y. Fang, M. Han, P. Ge, Z. Guo, X. Yu, Y. Deng, C. Wu, Q. Gong, and Y. Liu, "Photoelectronic mapping of the spin–orbit interaction of intense light fields," *Nat. Photonics* **15**(2), 115-120 (2021).

29. Y. Fang, Z. Guo, P. Ge, X. Ma, M. Han, X. Yu, Y. Deng, Q. Gong, and Y. Liu, "Strong-field photoionization of intense laser fields by controlling optical singularities," *Science China: Physics, Mechanics & Astronomy*, **64**(7), 274211 (2021).

30. A. Fleischer, O. Kfir, T. Diskin, P. Sidorenko, and O. Cohen, "Spin angular momentum and tunable polarization in high-harmonic generation," *Nat. Photonics* **8**(7), 543-549 (2014).

31. D. A. Kessler, and I. Freund, "Lissajous singularities," *Opt. Lett.* **28**(2), 111–113 (2003).

32. W. Miao, and G. Gbur, "Design of Lissajous beams," *Opt. Lett.* **47**(2), 297-300 (2022).

33. M. Lewenstein et al., "Theory of high-harmonic generation by low-frequency laser fields," *Phys. Rev. A* **49**(3), 2117 (1994).

34. N. Nagaosa, and Y. Tokura, "Topological properties and dynamics of magnetic skyrmions," *Nat. Nanotechnol.* **8**(12), 899−911, (2013).

35. X. Yi, Y. Liu, X. Ling, X. Zhou, Y. Ke, H. Luo, S. Wen, and D. Fan, "Hybrid-order Poincaré sphere," *Phys. Rev. A* **91**(2), 023801 (2015).

36. Y. Shen, et al., "Roadmap on spatiotemporal light fields," *arXiv preprint* arXiv:2210.11273 (2022).

37. Y. Fang, Z. Guo, P. Ge, Y. Dou, Y. Deng, Q. Gong, and Y. Liu. "Probing the orbital angular momentum of intense vortex pulses with strong-field ionization," *Light: Science & Applications* **11**(1), 34 (2022).

38. K. Litzius, et al., "Skyrmion Hall effect revealed by direct time-resolved X-ray microscopy," *Nat. Phys.* **13**(2), 170-175 (2017).

39. J. Miao, T. Ishikawa, I. K. Robinson, and M. M. Murnane, "Beyond crystallography: Diffractive imaging using coherent x-ray light sources," *Science* **348**(6234), 530-535 (2015).